\documentclass[article]{IEEEtran}
\IEEEoverridecommandlockouts
\usepackage{cite}
\usepackage{amsmath,amssymb,amsfonts}
\usepackage{algorithmic}
\usepackage{subcaption}
\usepackage{graphicx}
\usepackage{textcomp}
\usepackage{float}
\usepackage{xcolor}
\usepackage{tabularx}
\usepackage{array}
\usepackage{booktabs}
\usepackage{url}
\usepackage{changepage}

\begin{document}
\title{A Survey on Scheduling Techniques in the Edge Cloud: Issues, Challenges and Future Directions}

% Author Orchid ID: enter ID or remove command
\newcommand{\orcidauthorA}{0000-0002-8658-0098} % Add \orcidA{} behind the author's name
\newcommand{\orcidauthorB}{0000-0002-1288-7521} % Add \orcidB{} behind the author's name

% Authors, for the paper (add full first names)
\author{
\IEEEauthorblockN{Hassan Asghar and Eun-Sung Jung}\\
\IEEEauthorblockA{\textit{Department of Software and Communications Engineering} \\
\textit{Hongik University}\\
Sejong, South Korea \\
haxxanasghar@g.hongik.ac.kr, ejung@hongik.ac.kr}
}

% Current address and/or shared authorship
%\firstnote{Current address: Affiliation 3} 
%\secondnote{These authors contributed equally to this work.}
% The commands \thirdnote{} till \eighthnote{} are available for further notes

%\simplesumm{} % Simple summary

%\conference{} % An extended version of a conference paper

% Abstract (Do not insert blank lines, i.e. \\) 
\maketitle
\begin{abstract}
After the advent of the Internet of Things and 5G networks, edge computing became the center of attraction. The tasks demanding high computation are generally offloaded to the cloud since the edge is resource-limited. The Edge Cloud is a promising platform where the devices can offload delay-sensitive workloads. In this regard, scheduling holds great importance in offloading decisions in the Edge Cloud collaboration. The ultimate objectives of scheduling are the quality of experience, minimizing latency, and increasing performance. An abundance of efforts on scheduling has been done in the past. In this paper, we have surveyed proposed scheduling strategies in the context of edge cloud computing in various aspects such as advantages and demerits, QoS parameters, and fault tolerance. We have also surveyed such scheduling approaches to evaluate which one is feasible under what circumstances. We first classify all the algorithms into heuristic algorithms and meta-heuristics, and we subcategorize algorithms in each class further based on extracted attributes of algorithms. We hope that this survey will be very thoughtful in the development of new scheduling techniques. Issues, challenges, and future directions have also been examined.
\end{abstract}

% Keywords

% The fields PACS, MSC, and JEL may be left empty or commented out if not applicable
%\PACS{J0101}
%\MSC{}
%\JEL{}

%%%%%%%%%%%%%%%%%%%%%%%%%%%%%%%%%%%%%%%%%%
% Only for the journal Diversity
%\LSID{\url{http://}}

%%%%%%%%%%%%%%%%%%%%%%%%%%%%%%%%%%%%%%%%%%
% Only for the journal Applied Sciences:
%\featuredapplication{Authors are encouraged to provide a concise description of the specific application or a potential application of the work. This section is not mandatory.}
%%%%%%%%%%%%%%%%%%%%%%%%%%%%%%%%%%%%%%%%%%

%%%%%%%%%%%%%%%%%%%%%%%%%%%%%%%%%%%%%%%%%%
% Only for the journal Data:
%\dataset{DOI number or link to the deposited data set in cases where the data set is published or set to be published separately. If the data set is submitted and will be published as a supplement to this paper in the journal Data, this field will be filled by the editors of the journal. In this case, please make sure to submit the data set as a supplement when entering your manuscript into our manuscript editorial system.}

%\datasetlicense{license under which the data set is made available (CC0, CC-BY, CC-BY-SA, CC-BY-NC, etc.)}

%%%%%%%%%%%%%%%%%%%%%%%%%%%%%%%%%%%%%%%%%%
% Only for the journal Toxins
%\keycontribution{The breakthroughs or highlights of the manuscript. Authors can write one or two sentences to describe the most important part of the paper.}

%%%%%%%%%%%%%%%%%%%%%%%%%%%%%%%%%%%%%%%%%%
% Only for the journal Encyclopedia
%\encyclopediadef{Instead of the abstract}
%\entrylink{The Link to this entry published on the encyclopedia platform.}
%%%%%%%%%%%%%%%%%%%%%%%%%%%%%%%%%%%%%%%%%%

\section{Introduction}
Smart devices are gaining much attraction and playing an important role in augmented reality, NLP and facial recognition \cite{kumar_cloud_2010}. According to a CISCO report published in 2020 \cite{shehabi_united_2016}, states that there will be as many as 29,300 million network devices around the world till 2023. Likewise, there is a tremendous growth at the service providers side as well, due to the advancements in the telecommunication sector (5G technology) and Artificial Intelligence (deep learning algorithms) in the latest years. Most of the mobile devices are powerful enough and have multiple cores and some have GPUs as well for computation \cite{wu_machine_2019}.
\begin{figure}[h]
%\begin{adjustwidth}{-\extralength}{0cm}
  \centering
  \includegraphics[width=\columnwidth]{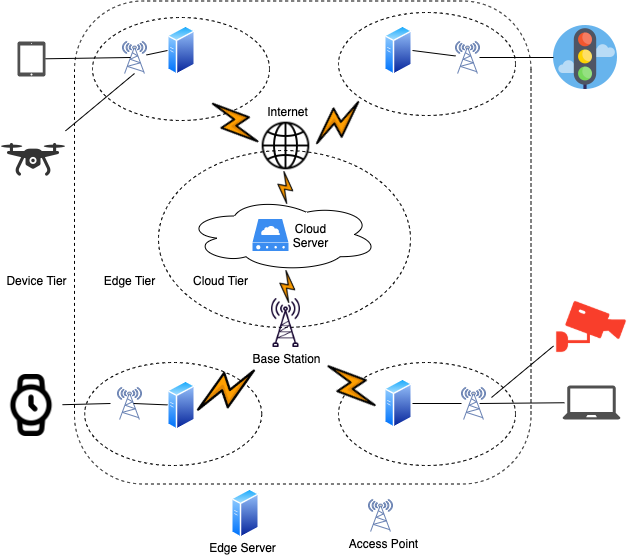}
  %\end{adjustwidth}
  \caption{The Edge Cloud computing scenario.}
  \label{tab:edgecloud}
\end{figure}
Despite all these advancements, there are some limitations due to bandwidth, power and size limits in mobile devices \cite{kumar_cloud_2010, ni_auction_2019}. Cloud computing is best suitable to cope with such hurdles and provision a smooth platform as it has an abundance of resources \cite{boukerche_sustainable_2019, qi_knowledge-driven_2019}. This type of approach does not apply to delay-sensitive applications like video analytics and real-time emergency preparedness. The above issue of delay sensitivity could be easily resolved by adopting the edge computing paradigm by moving the resources (computations) nearer to the mobile devices. But, the issue with edge computing \cite{hong_resource_2019} is its limited resources as compared to the cloud, therefore it cannot fulfil all the user's demand and Quality of Experience (QoE).

The Edge Cloud is a new paradigm that combines the benefits of cloud computing where resources are abundant and the edge computing that provision the delay-sensitive \cite{xu_task_2019} real-time applications but are resource-limited to solve all the problems mentioned above. Here, scheduling plays an important role in the efficiency of the Edge Cloud collaboration as your tasks should be scheduled either at the edge or at the cloud. There must be an efficient scheduling mechanism in the Edge Cloud that consider challenging issues like heterogeneous resources, inter-task dependence, user mobility and variable user requirements. The architecture of the Edge Cloud is given below in Figure~\ref{tab:edgecloud}. There are three tiers, the device tier, edge tier and cloud tier. In the device tier, every connected device computes its task locally and pushed to the cloud when the local resources are incapable to calculate.

Scheduling is the procedure of allocation of resources (computations) to the particular user that request service. In the Edge Cloud, this task is yet challenging to decide where to schedule the particular job at edge or cloud aiming to maximize the resource utilization. These days, there are multiple types of scheduling algorithms available like static, dynamic, mobility aware, machine learning based and other meta-heuristic algorithms. Dynamic allocation of resource, load balancing, reliability and availability are key challenges. Hence, there is always a need for state of the art scheduling algorithm techniques in the Edge Cloud environment.

In this paper, we have surveyed thoroughly to expose the state of the art scheduling techniques in the Edge Cloud. We have surveyed proposed scheduling strategies on merits and demerits, QoS parameters (energy consumption, SLA violations, resource utilization execution time, throughput, makespan time, availability, scalability and profit) along with diverse constraints such as deadline and fault-tolerance. In addition, various techniques and scheduling approaches have studied to see the impact that which one is feasible under certain circumstances and which one is to ignore. 

Shakarami et al. \cite{shakarami_review_2020}, Wang et al. \cite{wang_edge_2019} and Mach et al. \cite{mach_mobile_2017} improve the scheduling survey by classifying offloading techniques based on task types, user mobility, cooperativeness but none of the survey consider such a huge number of parameters simultaneously in the Edge Cloud collaboration. Still, there is a lot of insufficiency of futuristic research from a scheduling perspective. Therefore, we need a thorough state of the art survey that could integrate the research in scheduling. To the best of our knowledge, our survey paper shows the systematic and conceptual analysis of scheduling approaches and focuses on issues, challenges and future directions based on heuristics and meta-heuristics state of the art algorithms. Paper is organized into heuristic algorithms and meta-heuristics based on variable perspectives like dominance and drawback. We hope that this survey will be very thoughtful in the development of new scheduling techniques.

The rest of the paper is organized as follows: Section~\ref{tab:related_work} presents all the related work of various studies. Section~\ref{tab:methodology} then discusses the research methodology regarding the source of data and searching strategy. Section~\ref{tab:scheduling_problems} specifies the details of scheduling problems in Edge Cloud systems. Section~\ref{tab:categorization} describes the classification techniques of heuristic and meta-heuristic algorithms along with respective QoS parameters and varied constraints. This is followed by Section~\ref{tab:future} where the future directions are highlighted. At last, conclusion has been presented in Section~\ref{tab:Conclusion}.
\section{Related Work}\label{tab:related_work}
We are embracing the era of 5G communication, smart wearable and the Internet of Things (IoT). Smart mobile devices are getting popular day by day, the Edge Cloud is the promising paradigm to fulfil the variable user demands, especially for the delay-sensitive application requirements. Various work has been presented that examined the different surveys on resource allocation and scheduling in the Edge Cloud. The important point is that the Edge Cloud scheduling is still in the phase of infancy and lots of work needs to be explored. Here, we will see some survey papers that are related to the Edge Cloud scheduling and useful for our survey as well. Shakarami et al. \cite{shakarami_review_2020} presented a survey work on stochastic based scheduling where he categorized the task offloading mechanism into three fields: Markov chain, Markov process and hidden Markov. According to the best of our knowledge, this is the only work done in tasking scheduling techniques in the Edge Cloud perspective that has been presented by Wang et al. \cite{wang_edge_2019} where scheduling algorithms are categorized based on task types, user mobility, cooperativeness and multi-objective function like response time, energy consumption and load balancing is considered. Yang and Rahmani \cite{yang_task_2021} presented a literature review of fifteen articles related to variable task scheduling techniques based on two categories: heuristic and meta-heuristics. The advantages and limitations of task scheduling algorithms are explored. Mach and Becvar \cite{mach_mobile_2017} reviewed the edge computing architecture, where task offloading schemes, resource allocation and user mobility factor are taken into account based on QoS parameters.
In sum, the previous survey papers are lacking in the following points:
\begin{itemize}
    \item These survey papers did not account for the latest published article in connection with the Edge Cloud since we have considered the papers spanning 2015-2021.
    \item There is a lack of Machine Learning based scheduling technique's facet in the Edge Cloud scenario on the basis of advantage, demerits, Quality of Experience (QoS) and Quality of Experience (QoE) based.
    \item There is a lack of broader classification, taxonomy, and technical features of offloading mechanisms in the Edge Cloud.
    \item No paper evaluates such a massive number of parameters concurrently plus a detailed section concerning the Machine Learning aspect in the Edge Cloud.
\end{itemize}
\section{Research methodology: Data Source and Search Strategy}\label{tab:methodology}
The source of data that we have presented in this survey is from reputable journals and digital databases as given in the following Table~\ref{data}. We use the keywords as given in Table~\ref{search} to collect the related papers. We also use the logical operators AND and OR to connect multiple keywords to get more accurate results.
\begin{table}[H]
  \caption{Data Collection Source.\label{data}}
  \newcolumntype{C}{>{\centering\arraybackslash}X}
  \begin{tabularx}{\columnwidth}{ll}
    \toprule
    Publisher & Digital Library Link\\
    \midrule
ACM & \url{https://dl.acm.org} \\
IEEE & \url{https://ieeexplore.ieee.org} \\
Springer & \url{https://link.springer.com} \\
Hindawi & \url{https://www.hindawi.com} \\
Wiley Library &\url{https://onlinelibrary.wiley.com} \\ 
ResearchGate & \url{https://www.researchgate.net} \\
MDPI & \url{https://www.mdpi.com} \\
CNKI & \url{https://www.cnki.net} \\
ScienceDirect &\url{https://www.sciencedirect.com} \\
  \bottomrule
\end{tabularx}
\end{table}

\begin{table}[H]
  \caption{Data Searching Keywords.\label{search}}
  
  \newcolumntype{C}{>{\centering\arraybackslash}X}
  \begin{tabularx}{\columnwidth}{ll}
    \toprule
    Sr. No & Search Keywords\\
    \midrule
1 &  Edge Cloud \\
2 &  Scheduling \\
3 &  Resource allocation \\
4 &  Resource provisioning \\
5 &  Task offloading \\ 
6 &  Resource allocation \\
7 &  Scheduling algorithms \\
8 &  Scheduling techniques \\
9 &  Scheduling strategies \\
10 & Task scheduling \\
  \bottomrule
\end{tabularx}
\end{table}
All in all, we collected fifty (50) papers between the years 2015-2021 and focus on the latest development in the field of scheduling strategies in the Edge Cloud. We, therefore, collected papers that are published spanning over 2019-2021.
\section{Scheduling Problems In The Edge Cloud}\label{tab:scheduling_problems}
With the evolution of the Edge Cloud scenario, scheduling techniques are getting so much attraction in the recent past. In this section, we have examined different scheduling problems in Edge Cloud collaboration. Closest is a popular approach or principle and embraced widely in distributed systems as most User Equipments (UEs) needs to communicate with near edge cloud servers. For example, Han et al. \cite{han_ondisc_2019} conceded the latency in the edge cloud servers to minimize the total Weighted Response Time (WRT) to address the latency issue. Computation resources are equally distributed to all live jobs in line with their weights on the edge cloud servers by utilizing the concept of fairness. Unlike the closest principle, Wang et al. \cite{wang_reconciling_2016} investigated a task allocation problem and presented a two-phase based task scheduling model while improving the total cost considering the deadline. The task allocation model opts the low-cost cloudlets rather than choosing the closest edge server in the MEC environment.

Some studies conjecture that computation resources are abundant, and they only address the caching problem. A variety of solutions have been proposed based on request history to site the contents using caching services. Hou et al. \cite{hou_asymptotically_2016} addressed the caching problem by considering the switching cost. This approach is based on prior history to get the dynamic configuration in the Edge Cloud. Similarly, Li et al. \cite{li_neighborhood_2021} proposed a novel approach to solve the caching problem differently from the above solution based on the neighbourhood search concept that shortens the execution time, delay and workload allocation cost. In addition, their proposed approach was divided into four sub-problems: classification phase, resource allocation phase, clustering phase, and scheduling phase. Unlike, Huang et al. \cite{huang_task_2018} focused on the cache placement problem using a simulated annealing approach to mitigate the execution delay of the tasks in the Edge Cloud scenario. In addition, the model has partially trained using raw data at the edge to lessen the data transferring and preserve the privacy aspect. Additionally, Wei et al. \cite{wei_mobility-aware_2020} also resolve the content caching problem in the context of the Internet of Things (IoT) by considering that devices have limited computation capacity to reduce the transmission delay. Moreover, two algorithms are proposed as data allocation to operate suitable edge nodes, and the second one is data caching scheme to choose the appropriate cache function. In addition, Lyu et al. \cite{lyu_optimal_2017} also resolve the content caching and resource allocation problem in the context of cellular networks in the mobile edge computing (MEC) scenario. This study solved the caching problem differently from above as it exploits the alternate direction approach of multipliers based algorithm using partial knowledge results in maximization in system utility. In light of variance in user requests, Ma et al .\cite{ma_cooperative_2020} resolved the cooperative joint service placement and caching problem by proposing an iterative algorithm named ICE based on Gibbs sampling that shortens the response time in Mobile Edge Computing (MEC).

Substantial efforts have examined Mobile Edge Computing in the Internet of Things (IoT) context to mirror the practical conditions. Specifically, Hu et al. \cite{hu_dynamic_2020} presents a joint request offloading and resource scheduling problem to shorten the transmission delay in ultradense edge computing where an independent task model has been exploited to orchestrate User Equipments (UEs) requests. Likewise, Katsalis\cite{katsalis_sla-driven_2016} addressed joint offloading and scheduling problems via stochastic optimization to increase the profit while considering the SLA in time-critical systems. In addition, this model closely considers the scaling of VM's (e.g., number and type: small, medium and large) during the deployment period at every time step. In the same fashion, Li et al. \cite{li_cooperative_2021} proposed a heuristic approach to resolve the cooperative service placement and scheduling framework for cost optimization under varying user demands by considering the deadline constraint. Similarly, Yu et al. \cite{yu_energy-efficient_2018} derived a novel strategy that used a centralized approach to address the joint task unloading and resource scheduling problem in the Mobile Edge Cloud paradigm. Moreover, the proposed technique efficiently update the scheduling decision iteratively results in energy optimization and delay. Additionally, He et al. \cite {he_its_2018} presented a joint service placement and request strategy differently from above by introducing the concept of the shareable and non-shareable resources, aiming to increase the user requests. In addition, this study did not consider the timescale factor. Therefore, concurrent optimizing might increase the increase due to multiple updates at the Edge Cloud scenario.

Some studies pivoted on multi-layer architecture. For example, Huawei et al. \cite{huawei_adaptive_nodate} resolved the service provisioning problem for User Equipments (UEs) in the Edge Cloud environment. To increase the profit for network operators, a service update strategy is exploited while considering the access delay to enhance the traffic flow in the local cloud. In our point of view, this study is somewhat not feasible in real-time transmission scenarios as it ignores UAV's energy capacity and data replication factor. 

Other researchers focus on workflow scheduling while considering the uncertainties during the task execution and resolve the cost optimization problem in the hybrid cloud paradigm. Meng et al. \cite{meng_fault-tolerant_2019} proposed a solution based on reverse auction mechanism by incorporating a fault tolerance mechanism too, to shorten the system cost. Unsimilar, Wang et al. \cite{wang_learning_2018} proposed a Q-Learning based approach to tackle the same problem of uncertainties in the Edge Cloud (EC) environment. Moreover, the proposed strategy adopt mobility management through trial and error to mitigate the latency of the service. Similarly, Sun et al. \cite{sun_eco-friendly_2020} presented a two-step scheduling strategy to lessen the switching cost between edge and cloud servers where task arrival is uncertain. Next, Pollution Indicator Function (PIF) has been used to promote green energy, taking cost and latency tradeoff into account. Moreover, in the second phase of scheduling, Lyapunov based scheduling approach is proposed for the formulation of request dispatch problem that reduces cost and guarantees task latency. Besides, some researchers studied Meta-Heuristic based solutions for workflow scheduling problems in the Edge Cloud paradigm. For example, Xie et al. \cite{xie_novel_2019} suggested a Particle Swarm Optimization (PSO) based approach for task scheduling of a workflow, exploiting inertia weight update method and mutation operators. In addition, they also explored the tradeoff between makespan and processing cost. 

Apart from the above optimization objectives, Tuli et al. \cite{tuli_dynamic_2020} used the service migration concept for device placement and scheduling to minimize the latency and service level agreement in the Edge Cloud paradigm. Likewise, Urgaonkar et al. \cite{urgaonkar_dynamic_2015} also model service migration problem and takes the user mobility as an Markov Decision Process (MDP), also known as reinforcement learning to cut down the operational cost during the migration problem. Wang et al .\cite{wang_online_2017} addressed the same problem of service migration and proposed an online resource allocation model for cost optimization by considering the cloudlet's heterogeneity in the Edge Cloud. They presumed the workload that is supposed to calculate is "fluid" and divided into smaller chunks. We think this type of assumption does not hold any guarantee in real-world scenarios. Similarly, Miao et al. \cite{miao_intelligent_2020} presented a prediction based strategy exploiting LSTM, an extension of Recurrent Neural Networks to solve the task migration problem that shortens the latency in the Edge Cloud paradigm. On the other hand, this model is not feasible for the long term inter-task dependencies as the paper claimed. 

Several studies focused on the combination of C-RAN and MEC, where an intensive workload can be offload to the nearby edge cloud servers to save the battery life and compute capacity of the User Equipment (UE). Wang et al . \cite{wang_dynamic_2018} presented a framework for mobile service providers to address the network resource allocation problem. A time-slotted model is used to get the CPU cycle to calculate the running time of the task and then schedule it at the appropriate destination that increases the profit and power minimization. To avoid simultaneous migration problems, Farhadi et al. \cite{farhadi_service_2019} resolved the service placement and request scheduling problem under different optimization objectives (e.g., computation, communication and budget) for data-intensive services in Edge Cloud. Likewise, Alkhalaileh et al. \cite{alkhalaileh_data-intensive_2020} also resolved the resource allocation problem differently from the above model and proposed a task offloading scheme based on Mixed Integer Linear Programming (MILP) for joint optimization of cost and energy. Next, they solved the MILP problem by exploiting the Branch and Bound algorithm (BB), which results in higher complexity as problem size grow. Therefore, the presented model is more suitable for small and medium-size problems. Another thrust of research pivoted the radio resource allocation problem for heterogeneous mobile users at a commercial scale. For this, Fajardo et al. \cite{aguero_radio-aware_2015} proposed a novel Mobile Edge Scheduler for traffic flow in LTE downlink by deploying nearly at eNodeB. In addition, the proposed scheme is flow-aware and channel aware that shortens the mean delay in the LTE downlink. Moreover, this scheme is equally beneficial for network operators and third-party service providers. Unsimilar, Wang et al. \cite{wang_task_2020} proposed a novel strategy for task scheduling to the Edge Cloud paradigm by introducing a cataclysm strategy based on Catastrophic Genetic Algorithms that improve the makespan.

In the recent past, the Age of Information (AoI) has gained a lot of popularity and emerged as a new performance metric as it brings the concept of information freshness from the destination perspective. For example, Zhong et al. \cite{zhong_age-aware_2019} tailored a novel greedy scheduling approach to shorten the average AoI in the context of Edge Cloud collaboration. 

The Industrial Internet of Things (IIoT) is also gaining momentum in the research arena as it ushers the concept of smart cities, smart retail etc. For example, Kaur et al. \cite{kaur_keids_2020} presented a container scheduler named KEIDS to lessens the interference problem arise from other systems applications and improved energy utilization. The latest advancement in the sensors domain opened a new horizon for Wireless Sensor Networks (WSN) in the Internet of Things(IoT) perspective. Similarly, Ma et al. \cite{ma_study_2020} proposed a novel approach for the production system in an industrial set-up under the Edge Cloud environment. The presented distributed scheduling model is based on varying orders and for diverse requirements that guarantee real-time scheduling. 

Numerous studies have proposed workload scheduling problems in the Edge Cloud. For example, Wang et al. \cite{wang_computing_2019} resolved the dynamic resource allocation problem and suggested a support-vector-machine based scheme to improve the Quality of Experience (QoE) and bandwidth usage in Edge Cloud computing. Likewise, Sajnani et al. \cite{sajnani_latency_2018} resolved the workload allocation problem in the Edge Cloud differently and proposed a novel Multi-Layer Latency Aware Workload Assignment Strategy (MLAWAS) algorithm to shorten the job response time that first probe the best cloudlet then unloading the User Equipment (UEs) work onto that. In addition, this algorithm has sacrificed some latency to get a better response time.

In the modern era, applications are getting more complex, and even a simple system contains many interdependent tasks. To solve the problem of task allocation that are dependent on each other in multiple applications, Lee et al. \cite{lee_data_2020} model a scheduling algorithm named DATA to monitor the scheduling decisions and optimize the completion time and energy of Edge networks.

Every single server in Mobile Edge Computing (MEC) is enriched with limited compute power due to distributed deployment model. Therefore, executing a sheer amount of data on local servers might affect the performance in terms of QoS. To solve this, Liu et al. \cite{liu_cooper-sched_2019} presented a game based cooperative scheduling scheme named COOPER-SHED that is used to optimize the jobs in meeting guaranteed deadlines. 

The computing power in vehicles is not abundant, thus processing multiple tasks at once is nearly impossible, especially when an application is latency-sensitive. To solve this problem, Yin et al. \cite{yin_online_2017} derived a multi-resource allocation strategy that brings the computation capacity closer to the VANETs edge and assists the deadline hungry tasks to complete while fulfilling the Service Level Agreement (SLA) requirements.

Apart from resource management, load balancing is another major challenge of scheduling in a multiuser environment. In the cloud environment, there is a global view of available resources. Thus, it is easy to embrace a centralized approach. But, on the other hand, cloudlets are geographically dispersed, and there is no single central entity, resulting in cumbersome to effectuate load balancing. For example, Zhang et al. \cite{zhang_cost_2018} solved the load scheduling problem and proposed a two-stage algorithm that equally divides the load of the tasks to all the available nodes in the Edge Cloud environment. In addition to that, they improve the overall cost of the system while satisfying the Quality of Service (QoS) demands. Likewise, Zhao et al. \cite{zhao_load_2019} also solved the load imbalance scheduling problem differently by taking advantage of the fact that computation nodes have time-varying resource capabilities. The presented model is multi-objective optimization in nature that promote the resource allocation strategy by shortening the delay.Unsimilar to above, Lin et al. \cite{lin_distributed_2018} resolve the load balancing issue differently as they presented a sample-based model to address the load balancing issue that enhances the workload allocation performance and Quality of Experience (QoE) to users in the Edge Cloud paradigm. For this, it first checks edge resources before allowing another application. Li et al. \cite{li_efficient_2020} also solved the load scheduling problem by proposing a two-level scheduling approach based on the Artificial Fish Swarm Algorithm (AFSA) that shorten the latency, makespan and total cost.

The notable point is that Machine Learning (ML) and traditional computation offloading strategies are not contradictory but complement each other in Mobile Edge Computing (MEC) environment. Apart from the above solutions, many studies investigated the incorporation of Machine Learning (ML) techniques in computation offloading decisions in the Edge Cloud collaboration. For example, Crutcher et al. \cite{crutcher_hyperprofile-based_2017} resolved the server deployment problem differently through the k-Nearest Neighbor (kNN) algorithm based on historical data by augmenting Knowledge-Defined Networking (KDN). The objective of the derived strategy is to delay and energy utilization. Interestingly, the trained model is capable enough to accurately predict the system state, despite being trained on partial information. On the other hand, Meng et al. \cite{meng_delay-sensitive_2019} solved the task scheduling and unloading problem in User-Equipments (UEs) through implementing Deep Reinforcement Learning (DRL) based technique that reduces the average timeout and slowdown time stamp. Next, a reward-based function has been designed for the proposed algorithm to learn from the environment. Contrast with above, Zhao et al. \cite{zhao_selective_2019} solved the offloading problem differently as they proposed a selective offloading algorithm exploiting, ARIMA-BP approach. Next, devices that resided at the network's edge can consume less energy. Although the suggested model is pertinent regarding selective offloading, it encounters higher complexity.

In mission-critical applications, sometimes system parameters are not known in advance to learn the scheduling scheme. Therefore, it must have the capability to adopt that changes when encountering such a situation. Thus, to cope with this problem, Xu et al. \cite{xu_online_2017} presented a novel workload scheduling approach based on Reinforcement Learning (RL) by taking the battery capacity and green energy utilization into account. Similarly, Wang et al. \cite{wang_hidden_2017} resolved the resource offloading problem in the healthcare monitoring system and proposed a dynamic scheduling approach based on Hidden-Markov-Model (HMM) to lessen the latency and energy utilization. In addition, they analysed the tradeoff between latency and energy consumption thus suggested sacrificing some processing accuracy to improve the latency and adopted multiple scenarios for concept validation.

Augmented Reality (AR) has also gained an abundance of attraction in the recent past. But, it becomes more challenging when multiple users come into the picture in the AR game. For example, Jia et al. \cite{jia_delay-sensitive_2018} solved the multi-user game playing problem and proposed a decentralized user-coordination strategy to lessen the frame duration between players in the Edge Cloud paradigm. AR environment is interactive and demands ultra-low latency, if not provided, resulting in objects movement dizziness that hurt the Quality of Experience (QoE). Specifically, in a multi-user scenario, signal interference usually increases when users start offloading their computation simultaneously, results in channel quality degrades. To solve this problem, Chen et al .\cite{chen_efficient_2016} presented a distributed computation offloading scheme based on the game theory concept permitting User-Equipments (UEs) to make themself self-organized in the mobile Edge Cloud environment.

A plethora of research efforts have been studied microservices deployment problems to the Edge Cloud paradigm. Recently, Samanta et al. \cite{samanta_battle_2019} developed a novel microservice scheduling approach named FLAVOUR that improves multiple objectives (e.g., Quality of Service, Delay and Throughput). Similarly, Zhang et al. \cite{zhang_hetero-edge_2019} proposed a latency aware scheduling scheme to address the service placement problem at the microservices level using Apache Storm. First, the application is divided into storm tasks using Directed Acyclic Graph (DAG) and then mapped into respective heterogeneous resources (e.g., GPUs and CPUs) to lessen the End-to-End (E2E) latency.

Additionally, for the smooth running of operations, it is very crucial to attain the maximum profit. Hence, profit maximization is still a challenging issue in the Edge Cloud and demands some favourable attention. To this end, Wang et al. \cite{wang_incentive_2019} devised an incentive strategy to resolve this issue, exploiting a market-based pricing model that achieves profit maximization for Service Providers (SPs). Moreover, they adopt a static environment, but the user request arrival pattern is time-varying. Therefore, it seems nearly impossible to fetch this information in advance.
\section{Categorization of Scheduling techniques in the Edge Cloud}\label{tab:categorization}
\begin{figure*}
%\begin{adjustwidth}{-\extralength}{0cm}
\centering
\includegraphics[width=18cm, height=9cm]{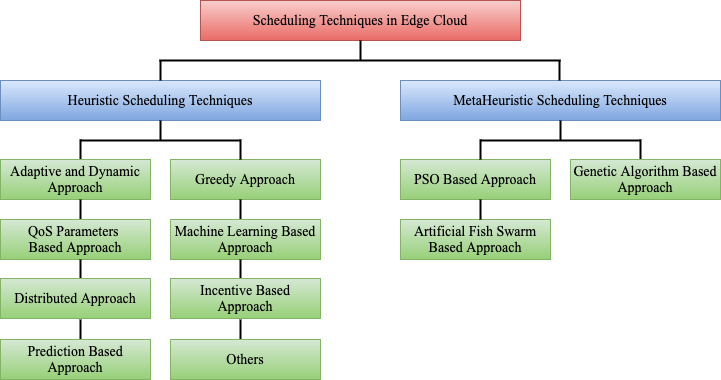}
%\end{adjustwidth}
\caption{Taxonomy of scheduling algorithms in the Edge Cloud.}    
\label{tab:edge_cloud_figure}
\end{figure*}

We have categorized our survey into two broader categories: heuristic and meta-heuristic algorithms. The taxonomy of scheduling algorithms in the Edge Cloud is presented in Figure~\ref{tab:edge_cloud_figure}. The ultimate goal of this survey is to provide a base for scheduling algorithms that leads towards novel scheduling algorithms.

\subsection{Heuristic Algorithms}
Heuristic algorithms are typically problem-specific and show excellent results for a particular field of problems. Since heuristic algorithms are domain-specific hence provide solutions within the fixed time period but are lacking in hard-optimization-problem solutions \url{https://en.wikipedia.org/wiki/Heuristic_(computer_science)}. In the Edge Cloud environment, numerous algorithms are presented, some of which consider workflow problems, some of which consider single task-level scheduling or application-level scheduling. We have segregated the heuristic algorithms based on multiple factors by analyzing the keywords from the article title and abstract. As a result, the categories in our taxonomy includes Adaptive and Dynamic Approach \cite{meng_fault-tolerant_2019,huawei_adaptive_nodate, hou_asymptotically_2016,hu_dynamic_2020,wang_dynamic_2018,tuli_dynamic_2020,urgaonkar_dynamic_2015,han_ondisc_2019,wang_online_2017}, Greedy Approach \cite{zhong_age-aware_2019,farhadi_service_2019}, QoS Parameters Based Approach \cite{zhang_cost_2018,lee_data_2020,yu_energy-efficient_2018,kaur_keids_2020,wang_reconciling_2016,katsalis_sla-driven_2016,wang_computing_2019,liu_cooper-sched_2019,li_cooperative_2021,yin_online_2017}, Machine Learning Based Approach \cite{meng_delay-sensitive_2019,huang_task_2018,zhao_selective_2019,crutcher_hyperprofile-based_2017,jia_delay-sensitive_2018,wang_hidden_2017,wang_learning_2018,xu_online_2017}, Distributed Approach \cite{lin_distributed_2018,sun_eco-friendly_2020,zhang_hetero-edge_2019,zhao_load_2019,li_neighborhood_2021,ma_study_2020,sajnani_latency_2018,samanta_battle_2019}, Incentive Based Approach \cite{wang_incentive_2019}, Prediction Based Approach \cite{miao_intelligent_2020,wei_mobility-aware_2020} and Others \cite{ma_cooperative_2020,alkhalaileh_data-intensive_2020,chen_efficient_2016,he_its_2018,aguero_radio-aware_2015,lyu_optimal_2017}. Review of different heuristic scheduling techniques together with dominance and drawbacks depicted in Table~\ref{heuristic_with_advantages_one},\ref{heuristic_with_advantages_two} and \ref{heuristic_with_advantages_three}.
\begin{table*}[htbp]
  \caption{Summary of Heuristic Scheduling Algorithms with Advantages and Limitations.\label{heuristic_with_advantages_one}}
  %\begin{adjustwidth}{-\extralength}{0cm}
		\newcolumntype{C}{>{\centering\arraybackslash}X}
		\begin{tabularx}{\textwidth}{p{.11\linewidth}p{.13\linewidth}p{.13\linewidth}p{.13\linewidth}p{.14\linewidth}p{.23\linewidth}}
    \toprule
    Paper (Year) & ML Technique & Tool Used & Metric &Advantages & Limitations\\
    \midrule
Meng et al. \cite{meng_fault-tolerant_2019} (2019) & Game theoretic algorithm & Workflow & PCP strategy, Reverse auction-based mechanism & Minimize the system cost with deadline satisfaction & Considers only one entry task and one exit task \\
%paper no. 14
Huawei et al. \cite{huawei_adaptive_nodate} (2017) & FFD, FFI, OFD, OFI & Independent & Arranging task in increasing and decreasing order & Considers profit, migration delay, and access delay & When server capacity is limited, only a small number of requests can be provisioned \\
%paper no. 15
Hou et al. \cite{hou_asymptotically_2016} (2016) & RL & Independent & RD and LRU policies have been used & Reduced the cost of the system & More performance parameters need to be addressed as an objective function \\
%paper no. 16
Hu et al. \cite{hu_dynamic_2020} (2020) & NCGG, MO-NSGA & Independent & NOMA based technique & Response time shortening and improved energy efficiency & When the resource of BS got full it degrades the performance and reduce response time, there are other bottlenecks as well in practical environment \\
%paper no. 17
Wang et al. \cite{wang_dynamic_2018} (2018) & VariedLen & Independent & Lyapunov based threshold policy & Improved the power efficiency and profit for MSP & Requests are queued in the buffer for a certain time period before dispatching increasing waiting time \\
%paper no. 18
Tuli et al. \cite{tuli_dynamic_2020} (2020) & Real time scheduler & Independent & A3C-R2N2 based policy & Shortened energy consumption, response time, SLA violation and cost & Performed better under fixed number of edge nodes and task, a little overhead would also occurs \\
%paper no. 19
Urgaonkar et al. \cite{urgaonkar_dynamic_2015} (2015) & Online control algorithm & Independent & Lyapunov optimization technique & Reduces operational cost & Solution is only efficient in other context if have same decoupling property \\
%paper no. 20
Han et al. \cite{han_ondisc_2019} (2019) & OnDisc & Independent & Speed augmentation model & Reduce the total weighted response time & Algorithm is clairvoyant as delay and processing time known in advance which is very tough in dynamic environment \\
%paper no. 21
Wang et al. \cite{wang_online_2017} (2017) & Online algorithm &	Independent	& Regularization technique & Cost reduction of service migration, service reconfiguration and delay & Capacity of resources is 2X times greater than the workload that leads towards underutilization of resources bearing extra cost \\
%paper no. 22
Zhong et al. \cite{zhong_age-aware_2019} (2019) & MIPR	&Independent	&Penalty function-based approach	&Reducing the overall age penalty of multiple end-users	&User could experience long waiting time when the traffic load is high \\
%paper no. 52
Fajardo et al. \cite{aguero_radio-aware_2015} & MESch&	Independent	&Flow-aware and channel-aware scheduling policy&	Optimizes the delivery of traffic flow and maximize cell throughput&	A coarse-grain CQI value is used during CG-CRR period it will degrade the performance in highly variable environments \\  
\bottomrule
\end{tabularx}
%\end{adjustwidth}
\end{table*}

\subsubsection{Adaptive and Dynamic Approach}
Adaptive scheduling techniques embrace changes at the run time based on prior criteria. Although in dynamic scheduling, information regarding the workload (task and node) is undisclosed, they frequently monitor the intended offloading node for load management in case of overutilization in the Edge Cloud environment. The following authors have studied scheduling based on adaptive and dynamic scheduling techniques. 

Meng et al. \cite{meng_fault-tolerant_2019} proposed a two-tier model for cost optimization and user-defined deadline satisfaction using Partial Critical Path (PCP) \cite{abrishami_deadline-constrained_2013} approach. First, they adopt a game-theoretic based scheduling mechanism to provide the optimal resources to user applications by employing the Vickrey–Clarke–Groves (VCG) \cite{lavi_algorithmic_2014} auction mechanism. Next, for mobility prediction of end devices random waypoint model \cite{kumar_performance_2015} has been used
\begin{table*}[htbp]
  \caption{Summary of Heuristic Scheduling Algorithms with Advantages and Limitations.\label{heuristic_with_advantages_two}}
  %\begin{adjustwidth}{-\extralength}{0cm}
		\newcolumntype{C}{>{\centering\arraybackslash}X}
		\begin{tabularx}{\textwidth}{p{.11\linewidth}p{.13\linewidth}p{.13\linewidth}p{.13\linewidth}p{.14\linewidth}p{.23\linewidth}}
    \toprule
    Paper (Year) & ML Technique & Tool Used & Metric &Advantages & Limitations\\
    \midrule
    Farhadi et al. \cite{farhadi_service_2019} (2019) &GSP-SS&	Independent	& Shadow scheduling technique	& Reduce operational cost under communication, computation, storage and budget constraints	&Separating placement and scheduling could bear extra network cost and overhead \\
%paper no. 24
Wang et al. \cite{wang_computing_2019} (2019) & EDF&	Independent&	Support vector based strategy	&Traffic classification and throughput improved&	Ingore the pre-emptive cost and arrival of tasks order is not considered \\
%paper no. 25
Liu et al. \cite{liu_cooper-sched_2019} (2019) & COOPER-SHED&	Independent	&EMA, Stable marriage game	&Addressed deadline guarantee issue by considering heterogeneous servers	&More QoS parameters are needed to incorporate \\
%paper no. 26
Li et al. \cite{li_cooperative_2021} (2021) & JCPS	&Independent&	Lyapunov based strategy&	Cost reduction and load balancing by exploiting spatial-temporal diversities in workload	&Have prior knowledge of all the resources that are being scheduled \\
%paper no. 27
Yin et al. \cite{yin_online_2017} (2017) & Online multiple resource allocation algorithm&	Independent	& Primal-dual approach& 	Maximize the profit by keeping in mind the SLA considerations and deadline constraints&	QoS service performance metrics needs to be incorporated in the proposed model \\
%paper no. 28
Zhang et al. \cite{zhang_cost_2018} (2018) & TTSCO&	Independent	&BF&	Cost reduction while QoS requirement satisfying&	Not compared with other state of the art algorithm to check efficiency \\
%paper no. 29
Lee et al. \cite{lee_data_2020} (2020) & DATA	&Independent	&Low-complexity based heuristic technique	&Effectively reduced the overall completion time of the tasks&	Container process only one task a the same time ,higher complexity \\
%paper no. 30
Yu et al. \cite{yu_energy-efficient_2018} (2018) & Distributed algorithm&	Independent	&Convex method adopted&	Energy efficiency improved along with completion time	&Algorithm works better only when tasks grow in size \\
%paper no. 31
Kaur et al. \cite{kaur_keids_2020} (2020) & KEIDS&	Independent&	Mosek solver method	&Improved energy efficiency and minimal interference&	Replicating more pods increases the network overhead \\
%paper no. 32
Wang et al. \cite{wang_reconciling_2016} (2016) & MCF-EDF	&Independent	&EDF policy	&Cost reducing by considering tight deadline guaranteed	&Complexity is high and there is a lack of comparing with state of the art heuristics \\
%paper no. 33
Katsalis et al. \cite{katsalis_sla-driven_2016} (2016) & LB VS& Independent &	Lyapunov optimization technique & 	Maximizes revenue by keeping less SLA's violations &	VM migration concept is lacking and load forecasting mechanism is not state of the art \\
Meng et al. \cite{meng_delay-sensitive_2019} (2019) & DRL&	Independent&	Reward-based strategy	&Shortened the period of slowdown tasks and better user experience&	Algorithm is better only for single-user as the user grows it increases the job completion time \\
%paper no. 35
Huang et al. \cite{huang_task_2018} (2018) & SAR algorithm &	Independent&	Probabilistic technique&	Reduced weighted transmission time and makespan by considering learning accuracy	&Scheduling algorithm is network topology dependent otherwise serious performance degradation in makespan \\
%paper no. 36
Lin et al. \cite{lin_distributed_2018} (2018) & DAA&	Independent&	Sampling-based load balancing technology&	Reduce the scheduling overhead and guaranteed quality of experience (QoE)&	Too many tasks cause a long waiting queue that affects the performance \\
\bottomrule
\end{tabularx}
%\end{adjustwidth}
\end{table*}

to improve the performance based on mobility. In addition, to cope with the delay faced during failure events, a fault tolerance strategy has derived. Experimental results showed that the proposed system outperforms in terms of cost and system delay. This study has used a single charge model for edge and cloud environments, which is different to the real world.

Huawei et al. \cite{huawei_adaptive_nodate} discussed a framework, named the Online-First (OF) algorithm, to augment the profit for service operators based on four heuristic algorithms as Online-First-Increasing, Online-First-Decreasing, First-Fit-Increasing and First-Fit-Decreasing. Online-First-Algorithm assigns priority to those mobile users who were first online. Next, the tasks grouped after arranging them in increasing and decreasing order using the algorithms (e.g., Online First Increasing (OFI) algorithm and Online First Decreasing (OFD) algorithm) respectively based on priority. Simulation depicted First Fit Increasing (FFI) and OFI posses greater profits, less access and migration time in contrast with rest. 

Hou et al. \cite{hou_asymptotically_2016} presented an intelligent configuration policy for tasks at the edge cloud as it has a great impact on the performance of the system. The proposed strategy is an online configuration named retrospective download with the least recently used (LRU). This policy has two-part the first one is RD and the second one is LRU. The functionality of RD is basically to judge either to download a service or not. The objective of the scheme is to reduce the total cost of the service including forwarding cost and the service download cost. Simulations were conducted with many offline algorithms using the real-world traces of data centres and proved it is asymptotically optimal and minimize the cost. 
Hu et al. \cite{hu_dynamic_2020} discussed an approch that solved the issue of response time and energy consumption in MEC environment. First, he formulated a power allocation (PA) problem and solve it using a non-cooperative model based on subgradient (NCGG). After that, a joint request offloading and resource scheduling problem has been formulated as MILP. To solve this a multi-objective algorithm has been proposed named MO-NSGA by implementing the NOMA technique. Simulations show that the proposed work outperforms as compared with the baseline algorithms. 

Wang et al. \cite{wang_dynamic_2018} described an extended work of Lyapunov based technique using multiple policies called VariedLen as Lyapunov critically assume that each job has an equal length in size. Three policies have been used in this proposed work i.e., a threshold policy for fronthaul links, a load balancing policy for the dispatching of jobs and a greedy policy for container scheduling. Simulation shows that the proposed work has near-optimal efficiency in profit as compared to competitive algorithms in MEC.
Tuli et al. \cite{tuli_dynamic_2020} explained a novel real-time scheduler based on Asynchronous-Advantage-Actor-Critic (A3C) learning having the capability to adapt to the dynamic changes of the environment and uses an Residual Recurrent Neural Network (R2N2) based architecture that considers hosts and tasks together. Experimental results show that the proposed model outperforms in terms of energy consumption by 14.4\%, response time by 7.74\%, SLA violation 31.9\% and cost by 4.64\% with a minimal scheduling overhead up to 0.002\%. 
Urgaonkar et al. \cite{urgaonkar_dynamic_2015} presented an online control algorithm to reduce the operational cost based on Lyapunov optimization. Mostly, this technique requires prior statistical knowledge, and the request arrival process is nonviable. To remove these complexities, the author proposed an online control algorithm that makes joint request for routing and configuration decisions, as a function of states then resultantly jointly decide service migration and workload scheduling. Simulations show promising results in terms of cost as compare to other such solutions. 

Han et al. \cite{han_ondisc_2019} proposed an online job dispatching and scheduling algorithm named OnDisc where every server at the Edge/Cloud follows the Higher Residual Density First (HRDF) rule to deal with unfinished jobs in the scheduling phase. Two jobs having the same (HRDF), the job which arrived earlier will be processed first. On the dispatching side, only those jobs will be dispatched first having/brings less increment to the total weighted response time. To bring fairness in the jobs, a fairness knob has also been introduced. Simulations based on real-world depicted OnDisc improved the total weighted response time as compared to baseline heuristic algorithms. As future work author suggested incorporating the network congestion as well. Wang et al. \cite{wang_online_2017} presented another online algorithm based on regularization technique. It simply divides the problem into multiple subproblems and then solved it in every independent sub-slots by taking the input from the previous slot produced. The online algorithm has no prior knowledge about user mobility and resource prices. Extensive experiment results proved that the proposed algorithm gained a 4x times reduction in total cost compared with the state of the art algorithms.
\subsubsection{Greedy Approach}
Greedy scheduling techniques typically focused on solving a problem based on the decision making which is best suitable for a specific situation \cite{malik_greedy_2013}.
The following authors have studied scheduling techniques using the greedy approach.

In recent times, the Age of Information has gained a lot of attraction through establishing a performance metric that apprises the freshness of status update information. Zhong et al. \cite{zhong_age-aware_2019} suggest a work conserving scheduling policy that reduces the weighted sum of Age of Information (AOI) over multiple users in the Edge-Cloud. In this greedy approach, scheduling updates the user requests and investigate the updated information. Meanwhile, the user having the maximum AOI will depict as an optimal policy. Simulation shows that it is a best work conserving strategy based on the age penalty function.
Farhadi et al. \cite{farhadi_service_2019} described a two-stage algorithm for cost reduction by using a greedy heuristic algorithm based on shadow scheduling. In the first phase, for request scheduling, an optimal algorithm is used under budget and communication constraints. Then they used a service placement algorithm in the second phase by combining the greedy heuristic with a service placement algorithm aiming to minimize the concurrent migrations. That gives a constant approximation ratio under variable constraints such as communication, computation, storage and budget.

\subsubsection{QoS Parameters Based Approach}
Various scheduling and/or resource provisioning strategies have proposed aiming to enhance the performance of QoS parameters (delay, execution time, makespan time, execution cost, resource utilization, energy, throughput, SLA violations, availability, scalability, profit etc.) to enhance the performance of the system.
Zhang et al. \cite{zhang_cost_2018} proposed a Best Fit (BF) based algorithm called Two-stage Task Scheduling Cost Optimization (TTSCO) that jointly reduces the cost while satisfying the delay requirements. It is a two-stage algorithm; in the first stage, it gets the preliminary scheduling strategy of all tasks using an improved version of the BF algorithm, and in the second stage, it revised the scheduling and got the final strategy for the scheduling of tasks. One drawback of this work is that it has not been compared with the state of the art algorithms.
Lee et al. \cite{lee_data_2020} proposed a heuristic algorithm called Dependency-Aware Task Allocation (DATA) to minimize the overall completion time of the application and energy utilization. The proposed work has three sub-algorithms; the first sub-algorithm creates a graph representing the sequence in which the requests have to be processed. The second sub-algorithm deals with the task allocation to the containers. And the last one does the scheduling of processing time of those tasks that were previously allocated. Simulation depicted that the proposed strategy reduces the overall completion time of the application compared with those algorithms that are dependency-unaware. 

Yu et al. \cite{yu_energy-efficient_2018} presented a distributed offloading algorithm based on the convex method to reduce energy efficiency and the completion time. All in all, it consists of offloading scheme selection, a configuration of clock frequency and power allocation control. Simulation depicted the proposed work as efficient in term of cost reduction up to 30\% as compared with the baseline algorithms.
Kaur et al. \cite{kaur_keids_2020} proposed a new approach to minimize the energy consumption named KEIDS based on Kubernetes as it is more scalable and lightweight as compared to traditional VMs. Precisely the proposed scheduler has two phases as “scheduling” and “synchronization”. In the former, the controller performs scheduling onto the available nodes aiming to minimize energy consumption. In the latter, the controller keeps an eye on the deployed containers to check, and do synchronization if

\begin{table*}[htbp]
  \caption{Summary of Heuristic Scheduling Algorithms with Advantages and Limitations.\label{heuristic_with_advantages_three}}
  %\begin{adjustwidth}{-\extralength}{0cm}
		\newcolumntype{C}{>{\centering\arraybackslash}X}
		\begin{tabularx}{\textwidth}{p{.11\linewidth}p{.13\linewidth}p{.13\linewidth}p{.13\linewidth}p{.14\linewidth}p{.23\linewidth}}
    \toprule
    Paper (Year) & ML Technique & Tool Used & Metric &Advantages & Limitations\\
    \midrule
    Sun et al. \cite{sun_eco-friendly_2020} (2020) & SCP&	Independent&	Lyapunov technique&	Impove the power consumption task dispatching	&Task dispatching to local edge servers is not included \\
%paper no. 38
Zhang et al. \cite{zhang_hetero-edge_2019} (2019) & Hetero-Edge&	Independent&	DAG based approach &	End-to-end latency reduction	&Edge cloud and traditional central cloud is not integrated if some want to use both models \\
%paper no. 39
Zhao et al. \cite{zhao_load_2019} (2019) & ETS	&Independent&	Augmented Lagrangian method& 	Successfully reduce the communication load&	There are dedicated servers for task computations that could lead to under utilization of the resources \\
%paper no. 40
Li et al. \cite{li_neighborhood_2021} (2021) & CANSS&	Independent&	K-nearest neighbor&	Effectively shortened the  execution time of data centers, CPU utilization and with improved memory&	Scale of the Edge Cloud collaboration is limited in proposed environment \\
%paper no. 41
Wang et al. \cite{wang_incentive_2019} (2019) & PMMRA &	Independent	&Matching matrix based&	Maximize the profit of resource providers&	Algorithm is not compared with other states of the art auction mechanisms \\
%paper no. 42
Miao et al. \cite{miao_intelligent_2020} (2020) & Task prediction algorithm&	Independent	&LSTM&	Efficiently reduce the  total task delay&	If there is some excessive load occur, the ongoing process will be halted \\
%paper no. 43
Wei et al. \cite{wei_mobility-aware_2020} (2020) & Service cache selection algorithm&	Independent	&Back propagation neural network	&Reduction in response time by considering the mobility factor of mobile users&	This algorithm is not efficient when user size grow there is serious degradation in response time \\
Samanta et al. \cite{samanta_battle_2019} (2019) & FLAVOUR	&Independent	&Policy based on SJF&	Novel and distributed technique to minimize the completion time and throughput	&Load balancing factor is ignored as it is important in micro-services and containerization app \\
%paper no. 45
Sajnani et al. \cite{sajnani_latency_2018} (2018) & MLAWAS	&Independent&	ETC matrix replica&	Earn a better response time&	Dynamic partitioning and load balancing needs to incorporated \\
%paper no. 46
Lyu et al. \cite{lyu_optimal_2017} (2017) & Asymptotically optimal offloading scheduler& 	Independent	&Lyapunov optimization technique& 	Maximization network utilization, balanced throughput and fairness &	Overall complexity of the algorithm is high and Deadline is not considered \\
%paper no. 47
Ma et al. \cite{ma_study_2020} (2020) & ECC &	Independent&	Prediction based strategy	& Periodically predict the total completion time of production tasks based on value added data&	Uncertainties  of real-time workflows system \\
%paper no. 48
Ma et al. \cite{ma_cooperative_2020} (2020) & ICE&	Independent	&Gibbs sampling technique&	Solve the issue of communication-computation trade-off and edge-node heterogeneity issue&	Task offloading causes additional transmission delay on LAN \\
%paper no. 49
Alkhalaileh et al. \cite{alkhalaileh_data-intensive_2020} (2020) & MILP&	Independent&	Mixed integer based strategy&	Minimize the monetary cost and device energy& 	Data files are stored in centralized local storage that is SPOF \\
%paper no. 50
Chen et al. \cite{chen_efficient_2016} (2016) & Distributed offloading algorithm	&Independent	&Game theoretic approach	&Superior computation offloading performance as user size grow&	Important performance metrics are missing \\
%paper no. 51
He et al. \cite{he_its_2018} (2018) & GSP-ORS, GSP GRS&	Independent&	Greedy approach	&Optimizes service placement and request scheduling&	Edge cloud can serve only if a user is a candidate and requested the resource otherwise not served \\

    \bottomrule
\end{tabularx}
%\end{adjustwidth}
\end{table*}
needed according to the optimization problem. It is worth notable here that the proposed scheduler also takes advantage of the \textit{etcd} module of Kubernetes which is the brain of the applications and store all the transactions. Simulation shows that the proposed scheduler outperforms in reducing energy consumption and interference due to the presence of other containers. Wang et al. \cite{wang_reconciling_2016} proposed an idea that span two phases. The main objective is to reduce the overall cost by considering the heterogeneity of cloudlets. The first phase is to select the cloudlet to decide where to offload the task for that an admission rate is used. The second phase is scheduling for a policy where Reverse-Task Scheduling (RTS) is used. RTS main goal is to decide the order of tasks to be executed so that the average time must be maximized. Finally, for scheduling, the proposed algorithm MCF-EDF comes into the picture and adopt the iterative process and scheduling the tasks by finding the critical interval \textit{I}* relative to its critical the Edge Cloud \textit{S}i*. Simulations are conducted in large and small scale environments and show that there is a promising gain in admission rate up to 30\% and cost reduction reasonably. As future work, the author left some work to explore the task portioning and allocation into related Edge Clouds.

Katsalis et al. \cite{katsalis_sla-driven_2016} introduced a Lyapunov optimization base algorithm named LB VS where the system has divided into multiple time slots. Controller enqueues all the service requests at the beginning of the slot. It is based on incentive and aiming to maximize the profit of the service providers. After that, there is a mapping service that map requests to VMs. For that, a greedy approximation algorithm has been adopted that sort the VMs in non-decreasing order relevant to VM type. Simulations depicted that the proposed strategy is better in terms of profit gaining by keeping fewer SLAs violations and being fair to service providers. The author intends to improve the load forecasting method along with incorporating VM migration. Deadline is a crucial aspect in the domain of scheduling since it is one of important quality of service (QoS) parameters. 
Wang et al.\cite{wang_computing_2019} proposed a computation aware scheduling in this work. The scheduling policy has two phases; in the first phase, SVM based multi-classifier has been adopted. In the second phase, a preemptive Earliest Deadline First (EDF) algorithm along with an admission control policy has been suggested. Numerical results depicted that it has better efficiency in terms of throughput than those without an admission control policy. How to retain the age of the classifier has been left as future work. 

Liu et al. \cite{liu_cooper-sched_2019} proposed a scheme called COOPER-SHED, a cooperative scheduler based on a stable marriage game to guarantee mobile users' deadline requirements. Both the mobile jobs and remote resources have been modelled on the cloudlet. Finally, the simulation shows the proposed algorithm has better efficiency and stability compared with the three heuristic algorithms FirstFit, MinCompletion and RoundRobin.
Li et al. \cite{li_cooperative_2021} discussed a joint co-operative and placement strategy by considering the resource cost and diversities in the workload. By taking the leverage of user deadline tolerance a Lyapunov based joint cooperative mechanism has been adopted for cost reduction and user satisfaction. 
Yin et al. \cite{yin_online_2017} 
presented an online resource allocation algorithm by taking into account the SLA and deadline factors. When a requested job arrives, the Edge Cloud makes the offloading decision without knowing prior assumptions and future arrivals. SLA properties are captured using the resource allocation problem. Every job is attached with a payment option that describes whether it makes a profit by execution at the Edge Cloud or not. The primal-dual approach has been applied that instantly takes the decision and informs the user about its job execution status.Results proved that the proposed work has a better competitive ratio as compared to the greedy approach.
\subsubsection{Machine Learning Based Approach} 
\begin{figure}[h]
%\begin{adjustwidth}{-\extralength}{0cm}
  \centering
  \includegraphics[width=\columnwidth]{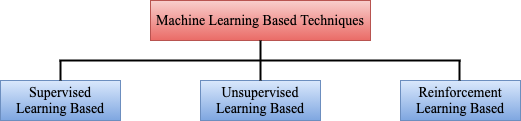}
  %\end{adjustwidth}
  \caption{Machine Learning based techniques taxonomy.}
  \label{tab:MLTaxonomy}
\end{figure}

\begin{figure}[h]
%\begin{adjustwidth}{-\extralength}{0cm}
  \centering
  \includegraphics[width=\columnwidth]{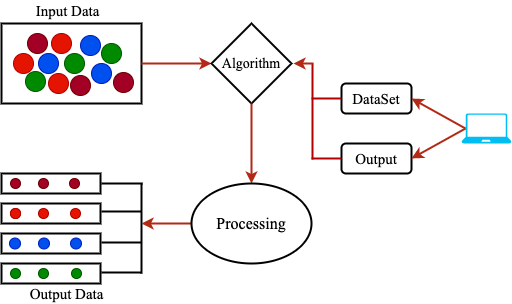}
  %\end{adjustwidth}
  \caption{Supervised learning model process.}
  \label{tab:Supervised}
\end{figure}

Over the last few years, Machine Learning is gaining much attraction since it can learn from input raw data by combining mathematics, statistics and machines to generate the output without any manual interruption \cite{lheureux_machine_2017}. We have classified the Machine Learning based scheduling techniques into three types; Supervised Learning, Unsupervised Learning and Reinforcement Learning. Based on the above literature, we have derived the following taxonomy presented in Figure~\ref{tab:MLTaxonomy}. The following authors have studied scheduling based on machine learning techniques.
\\
\textbf{Supervised Learning}: In Supervised Learning, there is an implied bonding between the input data and the data to predict. The training data set contain the input values (features) and the predicted output values. Supervised learning has human like analogy and learn from the past experience and improve the future predictions. The real time applications are speech and handwriting recognition \cite{liu_supervised_2011}. The working scenario of supervised learning is depicted in Figure~\ref{tab:Supervised}.

\begin{figure}[h]
%\begin{adjustwidth}{-\extralength}{0cm}
  \centering
  \includegraphics[width=\columnwidth]{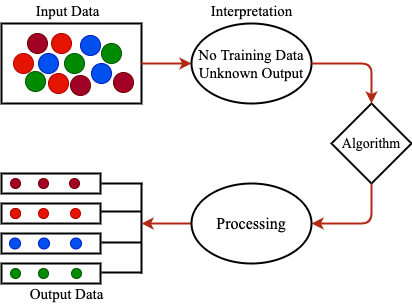}
  %\end{adjustwidth}
  \caption{Unsupervised learning model process.}
  \label{tab:Unsupervised}
\end{figure}

Zhao et al. \cite{zhao_selective_2019} proposed a technique named the ARIMA-BP-based Selective Offloading (ABSO) approach for energy minimization in mobile edge cloud computing while fulfilling the job delay constraint. A two-level level strategy has adopted. First, the ARIMA-BP model is used to predict the computation capability of the Edge Cloud. Second, A Selective Offloading Algorithm (SOA) has presented to get the offloading decision. Results depict a superior performance of the proposed strategy compared with the ABSO in reducing energy consumption. Crutcher et al. \cite{crutcher_hyperprofile-based_2017} presented a unique idea to resolve the offloading problem in edge cloud computing by leveraging the KDN approach for intelligent prediction regarding the offloading cost using the past data. First, it calculates the features for a {\bf Hyperprofile} based on the cost that has predicted previously. Finally, k-Nearest Neighbor (kNN) query has performed for node selection where a task is to be offloaded. In future work, authors want to incorporate more changes e.g; make a real-time testbed and enhance this Hyperprofile concept to routing and trust management.  \\
\textbf{Unsupervised Learning}: In unsupervised learning, the machine only acquires data input neither desired output nor any reward. In this approach, the machine aims at learning the representation that predicts the output. It could be useful in the mining patterns present in the data. Real-time application of unsupervised learning is clustering and dimensionality reduction \cite{bousquet_unsupervised_2004}. The workflow of Unsupervised Learning model is presented in Figure~\ref{tab:Unsupervised}.

\begin{figure}[h]
%\begin{adjustwidth}{-\extralength}{0cm}
  \centering
  \includegraphics[width=\columnwidth]{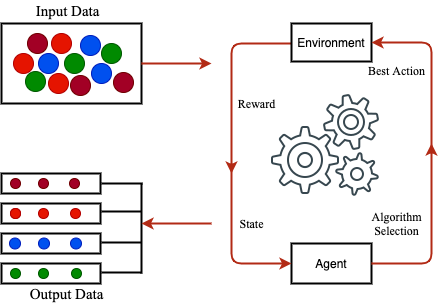}
  %\end{adjustwidth}
  \caption{Reinforcement learning model process.}
  \label{tab:Reinforcement}
\end{figure}

Jia et al. \cite{jia_delay-sensitive_2018} proposed a new approach aiming to reduce the game frame duration where there are multiple competing players in AR application. The problem has formulated as Decentralized Multiplayer Coordination (DMC). An iterative algorithm has proposed to solve the problem. Simulation shows that the proposed strategy will serve as a baseline in the near future AR multiplayer applications. As future work, the author wants to extend this work into stochastic traffic conditions. Wang et al. \cite{wang_hidden_2017} proposed a solution based on Hidden Markov Model (HMM) and incorporate an artificial neural network (ANN) consisting of a hidden layer and used thirty neurons. It will serve as a heartbeat classifier. three offloading strategies had used for scheduling named {\bf cloud-only}, {\bf hybrid-cloud} and {\bf mobile-only}. Energy and delay are considered as performance evaluation parameters. Simulation proved that the proposed work identify the better configuration.
\\
\textbf{Reinforcement Learning}: In Reinforcement Learning, a machine interact with its environment and produce actions. Next, these generated actions may change the state and receive a reward in turn. The ultimate goal is to maximize the rewards and reduce the punishment throughout the lifespan \cite{bousquet_unsupervised_2004}.
The workflow of Reinforcement Learning model is shown in Figure~\ref{tab:Reinforcement}.

Wang et al. \cite{wang_learning_2018} consider the factor of mobility and proposed an approach based on Q-Learning to address the problem of uncertainty in the Edge Cloud. The environment learns from trial and error strategy and then learns optimal management for delay reduction. The state is maintained and Q-Values are updated regularly. Next, the offloading decision will take place based on the highest Q-Value. Meng et al. \cite{meng_delay-sensitive_2019} has proposed a learning-based user policy an algorithm named Deep Reinforcement Learning (DRL) that convert the optimization problem into a learning problem. A reward-based function has also been introduced that learn the offloading mechanism after a period of training. This algorithm works simultaneously for scheduling and resource allocation. Simulation depicted that it has better efficacy as compared to the FCFS, SJF and Random in terms of average slowdown period and average timeout period. 
Huang et al. \cite{huang_task_2018}, described two approaches to reduce the weighted transmission time. The first technique is the extreme point LP Solution based algorithm and the second one is the Simulation Annealing rearranging algorithm which reduced the makespan time. The edge learning framework consists of two steps, the scheduling process and the deep learning process. In the former, mobile users send the information of data sizes and then select the relevant deep learning model. After that, the cloud collects the information followed by the execution of the scheduler and send back data to edge servers. In the latter, the training data is generated by mobile users. Simulations show there is a great reduction in makespan and weighted transmission time. Xu et al. \cite{xu_online_2017} presents a scheme namely post-decision state (PDS) which consist of offline and online phase. Value iterations have performed in the offline phase and in the online stage, there is an RL that learns the best policy for scheduling decision. For joint offloading author proposed an MDP structure that has also incorporated in the PDS. Simulation shows a significant improvement in Edge Cloud scheduling. Machine Learning based scheduling approaches along with advantages, limitations, tool used and performance metrics are given in Table~\ref{ML_Advan_limitation}.
\begin{table*}[htbp]
  \caption{Machine Learning Based Scheduling Techniques with Advantages and Limitations.\label{ML_Advan_limitation}}
  %\begin{adjustwidth}{-\extralength}{0cm}
		\newcolumntype{C}{>{\centering\arraybackslash}X}
		\begin{tabularx}{\textwidth}{p{.11\linewidth}p{.13\linewidth}p{.13\linewidth}p{.13\linewidth}p{.14\linewidth}p{.23\linewidth}}
    \toprule
    Paper (Year) & ML Technique & Tool Used & Metric &Advantages & Limitations\\
    \midrule
Zhao et al. \cite{zhao_selective_2019} (2019) & Deep Neural Network (DNN) & Simulation& Energy, Delay & Selective task offloading & Overall complexity is high \\
Crutcher et al. \cite{crutcher_hyperprofile-based_2017} (2017) & K-Nearest Neighbors (KNN)& Simulation & Energy, Delay & Easy implementation &Take time to learn\\
Jia et al. \cite{jia_delay-sensitive_2018} (2018) & Game AI & Simulation & Delay & Reducing the game frame duration when there are multiple players & System model is too complex to adopt \\
Wang et al. \cite{wang_hidden_2017} (2017) & Hidden Markov Model & Real & Energy, Delay &  Reducing the processing latency and energy consumption & Technical aspects are hidden and organisation is weak 
\\
Wang et al. \cite{wang_learning_2018} (2018) & Q-Learning & Simulation & Delay, Energy & Simplicity of implementation & Very low convergence rate and state of the art comparison is lagging\\
Xu et al. \cite{xu_online_2017} (2017) & Online Learning & Simulation & Cost and energy consumption & Leveraging the green energy concept & Load balancing have not been considered \\
\bottomrule
\end{tabularx}
%\end{adjustwidth}
\end{table*}

\subsubsection{Distributed Approach} In resource provisioning, resource allocation and scheduling done using different environments called distributed scheduling techniques. The following authors have studied scheduling based on distributed scheduling approach. Lin et al. \cite{lin_distributed_2018} explained a distributed scheduling strategy named distributed and application-aware algorithm (DAA) based on sampling load based technology. The author describes the scheduling policy as a two-fold challenge firstly as the Edge Clouds are distributed geographically secondly the task that is supposed to be offloaded into the Edge Cloud are of different types. Under such situations to provide quality of experience (QoE) to the user is quite challenging. A greedy approach has been adopted to solve this issue. Simulation shows that the proposed scheduling policy has better performance in terms of quality of experience and improved makespan time. Sun et al. \cite{sun_eco-friendly_2020} proposed a new approach to make the environment clean along with guaranteed delay sensitivity of task dispatching. A Sequential Convex Programming (SCP) algorithm with a pollution indicator function has adopted for the pollution measurement and suggest using cleaner power. For task dispatching,  a Lyapunov optimization-based distributed task scheduling mechanism is used. Simulation depicted that it effectively improve the usage of clear power up to 50-60\%. In future work, the author is eager to use some prediction strategy for unit cost prediction of power storage.
Zhang et al. \cite{zhang_hetero-edge_2019} proposed a latency-aware scheme named Hetero-Edge based on DAG strategy and built on Apache Storm, consists of multiple heterogeneous computing resources like CPUs and GPUs. The idea behind Hetero-Edge is that it divides the application into Storm tasks which are defined by DAG and then send to the heterogeneous edge servers for speedy processing. Simulation proves that the proposed strategy successfully gained improvement in end-to-end latency up to 40\% compared with storm scheduling.
Ma et al. \cite{ma_study_2020} proposed a novel Edge Cloud collaborative scheduling framework to cope with the dynamic fluctuation in a factory environment. The proposed model is divided into two parts as headquarters cloud side prediction model and edge-side scheduling optimization model. In the former, there are three types of data involved based on which the prediction is made that will be helpful in another module. After prediction, the data of customer order is further divided into sub-tasks to schedule at the Edge Cloud. Example analysis proved the rationality of the work. As a future work author intend to explore some diversified verification method.

Zhao et al. \cite{zhao_load_2019} presented a coded MapReduce based algorithm  named ETS aiming to reduce the load scheduling problem. The framework is divided into two parts in an offline and online setting. In the former, an augmented Lagrangian method is used to get the computation load and in the latter, the worst-case performance bound is derived using the offline setting. Past information is used for planning and learning-based ETS is used for load scheduling optimally. Simulation proved it has superior performance as compared with greedy, capacity and fair schedulers. Li et al. \cite{li_neighborhood_2021} suggested a cache-aware algorithm based on neighbourhood search mechanism as classification, allocation of resources, clustering of node and finally the main algorithm. In the job classification phase, jobs are divided into three levels as level1, level2 and level3 based on Naïve Bayes theory. In clustering, nodes having similar ability are grouped using neighbourhood search and then cached by delay-waiting. Jobs that are not promising the demand of data locality needs to be offloaded to the nodes having similar capacity. Simulations proved that there is a great reduction in delayed content transmission and placement along with shortened execution time. As a future work author want to incorporate AI algorithms for the prediction of either caching locality that could be achieved in waiting time.

Sajnani et al. \cite{sajnani_latency_2018} presented a Multi-Layer Latency Aware Workload Assignment strategy (MLAWAS) to offload the workload into the optimal cloudlet. This strategy first accepts the workload that is coarse-grained and adopted the Poisson process. Before actual offloading took place, it first schedules the optimal network via it is supposed to send and the information on that path is supposed to be known in advance. After network selection, optimization of the communication delay will be performed along with a learning factor that takes the decision and decide the score for offloading and the task would be allocated to the best optimal cloudlet. Simulation depicted there is a promising reduction in terms of response time has earned.
Samanta et al. \cite{samanta_battle_2019} proposed a novel and distributed technique, namely FLAVOUR for tasks scheduling. In the design of FLAVOUR, the author first established a stochastic service delay minimization problem by considering completion time and network stability as constraints. After solving this, a latency aware optimal scheduling problem will establish the theoretical foundation for FLAVOUR. Simulation shows FLAVOUR outperforms in terms of latency and throughput.
\subsubsection{Incentive Based Approach} Incentive-based scheduling techniques provide an equal ground for mobile users and Service Providers (SP) to play. From the perspective of the Mobile User (MU), given that there are two jobs having identical lengths, priority should be given to that particular job that pays more and must encounter a better quality of experience in comparison with others who pay less. Here, quality of experience (QoE) means shorter response time and ultra-low latency \cite{zhu_incentive-based_2006}.
Wang et al. \cite{wang_incentive_2019} proposed a Profit Maximization Multi-Round Auction (PMMRA) strategy to increase the profit of the Service Providers (SP) in a market competitive based model. Mobile users submit their demand to the auctioneer and SP submit the bid information relative to the information provided by the auctioneer. When bidding complete three tasks need to be performed by the auctioneer which is user matching, check computational resources and estimation of utilities. After that, matrix determination will decide the final price that best for the user. Simulation proved that the proposed works better and maximizes the profit for the service provider.
\subsubsection{Prediction Based Approach}
In order to effectively utilizing the resource in the Edge Cloud environment, it is need of the hour to have an effective prediction based scheduling approach aiming to maximize the performance and reduced execution time together with delay sensitivity \cite{kaur_prediction_2021}. Therefore, it's vital to predict the requirement of the resource and then scheduled them into the edge/cloud location. The following authors have studied scheduling based on prediction strategy. 
Miao et al. \cite{miao_intelligent_2020} proposed a prediction based algorithm jointly using the capabilities of Artificial Intelligence (AI) to reduce the total task delay. Briefly, the task is predicted using the LSTM strategy and the predicted data will be used to offloading the task to the Edge Cloud. Here, the main goal is to find as much accuracy as one can because this predicted data will be used later on for task offloading decision. In offloading, there are three approaches to consider either the task needs to be processed locally or at the edge cloud or partially locally and the remaining part at the edge cloud. Simulation depicted that the proposed strategy works very efficiently and reduce the total task delay.
Wei et al. \cite{wei_mobility-aware_2020} proposed a service cache selection algorithm based on a backpropagation neural network to solve the service caching issue by considering the factor of mobility of Mobile Users (MUs). First, it will predict the location of the user by adopting the geometric model. Once a prediction is completed the role of the service allocation algorithm begins. Here, there are two scenarios as if the user falls in the overlapping area the request will be forwarded to the base station where services are cached otherwise proportion of target is calculated and the base station having a larger probability will be served. Simulation shows the proposed work has improved response time by considering mobility factor and great reduction in request failure. In future work, the author wants to optimize the location prediction when there is no GPS information. In addition to that, the dataset will be improved to train the model on real data.
\subsubsection{Others}In recent years, there are some other approaches as well for scheduling in the Edge Cloud.
Ma et al. \cite{ma_cooperative_2020} proposed an iterative algorithm called ICE to solve the issue of coupling subproblem, communication-computation tradeoff and edge-node heterogeneity. The algorithm consisted of two layers, the inner layer and the outer layer. The outer layer updates the caching policy iteratively and based on Gibbs sampling while the inner layer performed the scheduling of workload based on the idea of water filling. Simulations proved that the proposed algorithm reduces the service response time and traffic offloading issue efficiently as compared to the benchmark algorithms.
Alkhalaileh et al. \cite{alkhalaileh_data-intensive_2020} proposed a MILP based strategy to reduce the monetary cost and device energy as an objective for an optimization problem. The proposed algorithm considers variable deadlines and multi-user parameters into account. Simulation shows that it has better efficiency as compared with Particle Swarm Optimization (PSO).

Lyu et al. \cite{lyu_optimal_2017} presented, a asymptotically optimal scheduling model based on Lyapunov technique. A threshold value has been derived having partial out-of-date information. IoT devices meeting the threshold will be designed for feedback, ultimately reducing it asymptotically. Extensive simulations show that it outperforms as only 60 out of 5000 going to send feedback as this decline is due to an increase in the number of devices.

Chen et al. \cite{chen_efficient_2016} proposed a distributed computation offloading algorithm to solve the multi-user offloading issue as a multi-user game. It first achieves the Nash equilibrium then derive the upper bound and quantify the ratio in term of performance metrics. Simulation corroborates that the proposed scheme has the upper hand as compared to the state of the art algorithms.
He et al. \cite{he_its_2018} presented a constant-factor approximation algorithm for jointly service placement and request scheduling by considering hard constraints of communication and computation. Two types of decision variables are introduced for decision making which indicates either the service l should dispatch at the Edge Cloud n, secondly, a user u should schedule onto the Edge Cloud n or not. To fulfil these two, ILP has been formed and GSP-ORS and GSP-GRS have been used the difference is only the
\begin{table*}[htbp]
\caption{Summary of Heuristic Scheduling Algorithms with QoS Parameters.\label{heuristic_qos_one}}
%\begin{adjustwidth}{-\extralength}{0cm}
		\newcolumntype{C}{>{\centering\arraybackslash}X}
		\begin{tabularx}{\textwidth}{p{.19\linewidth}p{.05\linewidth}p{.05\linewidth}p{.05\linewidth}p{.05\linewidth}p{.07\linewidth}p{.04\linewidth}p{.07\linewidth}p{.02\linewidth}p{.07\linewidth}p{.06\linewidth}p{.05\linewidth}}
			\toprule
Paper (Year) &	Execution Time&	Makespan Time&	Execution Cost&	Response Time&	Resource Utilization	&Energy&	Throughput&	SLA &	Availability&	Scalability	&Profit\\ 
\midrule
Meng et al. \cite{meng_fault-tolerant_2019} (2019) & x	&x	&o	&x	&x	&x	&x	&x	&x	&o	&x \\
Huawei et al. \cite{huawei_adaptive_nodate} (2017) & x	&x	&x	&x	&x	&x	&x	&x	&x	&x	&o \\
Hou et al. \cite{hou_asymptotically_2016} (2016) & x	&x	&o	&x	&x	&x	&x	&x	&x	&x	&x \\
Hu et al. \cite{hu_dynamic_2020} (2020) & x	&x	&x	&o	&x	&o	&x	&x	&x	&x	&x \\
Wang et al. \cite{wang_dynamic_2018} (2018) & x	&x	&x	&x	&x	&o	&x	&x	&x	&x	&o  \\
Tuli et al. \cite{tuli_dynamic_2020} (2020) & x	&x	&o	&o	&x	&o	&x	&o	&x	&x	&x \\
Urgaonkar et al. \cite{urgaonkar_dynamic_2015} (2015) & x	&x	&o	&x	&x	&x	&x	&x	&x	&x	&x \\
Han et al. \cite{han_ondisc_2019} (2019) & x	&x	&x	&o	&x	&x	&x	&x	&x	&x	&x \\
Wang et al. \cite{wang_online_2017} (2017) & x	&x	&o	&x	&x	&x	&x	&x	&x	&x	&x \\
Zhong et al. \cite{zhong_age-aware_2019} (2019) & x	&x	&x	&x	&o	&x	&x	&x	&x	&x	&x \\
Farhadi et al. \cite{farhadi_service_2019} (2019) & x	&x	&o	&x	&x	&x	&x	&x	&x	&x	&x \\
Wang et al. \cite{wang_computing_2019} (2019) & x	&x	&x	&x	&x	&x	&o	&x	&x	&x	&x \\
Liu et al. \cite{liu_cooper-sched_2019} (2019) & x	&x	&x	&x	&o	&x	&x	&x	&x	&x	&x \\
Li et al. \cite{li_cooperative_2021} (2021) & x	&x	&o	&x	&o	&x	&x	&x	&x	&x	&x \\
Yin et al. \cite{yin_online_2017} (2017) & x	&x	&x	&x	&x	&x	&x	&o	&x	&x	&o \\
Zhang et al. \cite{zhang_cost_2018} (2018) & x	&x	&o	&x	&o	&x	&x	&x	&x	&x	&x \\
Lee et al. \cite{lee_data_2020} (2020) & o	&x	&x	&x	&x	&x	&x	&x	&x	&x	&x \\
Yu et al. \cite{yu_energy-efficient_2018} (2018) & o	&x	&x	&x	&x	&o	&x	&x	&x	&x	&x \\
Kaur et al. \cite{kaur_keids_2020} (2020) & x	&x	&x	&x	&o	&o	&x	&x	&o	&o	&x \\
Wang et al. \cite{wang_reconciling_2016} (2016) & x	&x	&o	&x	&x	&x	&x	&x	&x	&x	&x \\
Katsalis et al. \cite{katsalis_sla-driven_2016} (2016) & x	&x	&x	&x	&x	&x	&x	&o	&x	&x	&o \\
Meng et al. \cite{meng_delay-sensitive_2019} (2019) & o	&x	&x	&x	&x	&x	&x	&x	&x	&x	&x \\
Huang et al. \cite{huang_task_2018} (2018) & o	&o	&x	&x	&x	&x	&x	&x	&x	&x	&x \\
Lin et al. \cite{lin_distributed_2018} (2018) & x	&o	&x	&x	&x	&x	&x	&x	&x	&x	&x \\
Sun et al. \cite{sun_eco-friendly_2020} (2020) & x	&x	&x	&x	&x	&o	&x	&x	&x	&x	&x \\
Zhang et al. \cite{zhang_hetero-edge_2019} (2019) & o	&x	&x	&x	&x	&x	&x	&x	&x	&x	&x \\
Zhao et al. \cite{zhao_load_2019} (2019) & x	&x	&x	&x	&o	&x	&x	&x	&x	&x	&x \\
Li et al. \cite{li_neighborhood_2021} (2021) & o	&x	&o	&x	&o	&x	&x	&x	&x	&x	&x \\
Wang et al. \cite{wang_incentive_2019} (2019) & x	&x	&x	&x	&x	&x	&x	&x	&x	&x	&o \\
Miao et al. \cite{miao_intelligent_2020} (2020) & o	&x	&x	&x	&x	&x	&x	&x	&x	&x	&x \\
Wei et al. \cite{wei_mobility-aware_2020} (2020) & x	&x	&x	&o	&x	&x	&x	&x	&x	&x	&x \\
Samanta et al. \cite{samanta_battle_2019} (2019) & o	&x	&x	&x	&x	&x	&o	&x	&x	&x	&x \\
Sajnani et al. \cite{sajnani_latency_2018} (2018) & x	&x	&x	&o	&x	&x	&x	&x	&x	&x	&x \\
Lyu et al. \cite{lyu_optimal_2017} (2017) & x	&x	&x	&x	&o	&x	&o	&x	&x	&x	&x \\
Ma et al. \cite{ma_study_2020} (2020) & o	&x	&x	&x	&x	&x	&x	&x	&x	&x	&x \\
Ma et al. \cite{ma_cooperative_2020} (2020) & x	&x	&x	&o	&x	&x	&x	&x	&x	&x	&x \\
Alkhalaileh et al. \cite{alkhalaileh_data-intensive_2020} (2020) & o	&x	&o	&x	&x	&o	&x	&x	&x	&x	&x \\
Chen et al. \cite{chen_efficient_2016} (2016) & x	&x	&x	&x	&x	&x	&x	&x	&x	&o	&x \\
He et al. \cite{he_its_2018} (2018) & o	&x	&x	&x	&x	&x	&x	&x	&x	&x	&x \\
Fajardo et al. \cite{aguero_radio-aware_2015} & x	&x	&x	&x	&x	&x	&o	&x	&x	&x	&x  \\
\bottomrule
\end{tabularx}
%\end{adjustwidth}
\end{table*}

\begin{table*}[htbp]
\caption{Summary of Meta-Heuristic Scheduling Algorithms with Advantages and Limitations.\label{metaheuristic_advantages_one}}
%\begin{adjustwidth}{-\extralength}{0cm}
		\newcolumntype{C}{>{\centering\arraybackslash}X}
		
\begin{tabularx}{\textwidth}{p{.11\linewidth}p{.13\linewidth}p{.11\linewidth}p{.13\linewidth}p{.15\linewidth}p{.24\linewidth}}
\toprule
Paper (Year)  & Algorithm & Task & Technique & Advantages & Limitations\\
\midrule
%paper no. 53
Xie et al. \cite{xie_novel_2019} (2019) & DNCPSO&	Workflow&	Non-linear inertia weight with selection and mutations operations&	Maintain the trade-off between total execution time and economic cost	&Easy to fall into local optimum and difficult to obtain real optimal solution \\
%paper no. 54
Li et al. \cite{li_efficient_2020} (2020) & MIAFSA&	Independent&	Artificial fish swarm based, considering load balancing&	Decreasing completion time and total cost	&Assuming hard dealines is not favourable when there are multiple task \\
%paper no. 55
Wang et al. \cite{wang_task_2020} (2020) & CGA	&Independent&	Roulette selection strategy, optimization mutation and crossover operator &	Shortening the task completion basis on satisfying task delays, global optimization&	Inter-dependencies of the tasks are not considered \\
\bottomrule
\end{tabularx}
%\end{adjustwidth}
\end{table*}

\begin{table*}[htbp]
\caption{Summary of Meta-Heuristic Scheduling Algorithms with QoS parameters.\label{metaheuristic_qos_one}}
%\begin{adjustwidth}{-\extralength}{0cm}
		\newcolumntype{C}{>{\centering\arraybackslash}X}
\begin{tabularx}{\textwidth}{p{.19\linewidth}p{.05\linewidth}p{.05\linewidth}p{.05\linewidth}p{.05\linewidth}p{.07\linewidth}p{.04\linewidth}p{.07\linewidth}p{.02\linewidth}p{.07\linewidth}p{.06\linewidth}p{.05\linewidth}}
\toprule
Paper (Year)&	Execution Time &	Makespan Time&	Execution Cost&	Response Time&	Resource Utilization	&Energy&	Throughput&	SLA &	Availability&	Scalability	&Profit\\ 
\midrule
Xie et al. \cite{xie_novel_2019} (2019) &

    x&	o&	o&	x&	x&	x&	x&	x&	x&	x&	x  \\
Li et al. \cite{li_efficient_2020} (2020) & o&	x	&o&	x&x&x&x&	x&	x	&x&	x \\
Wang et al. \cite{wang_task_2020} (2020) & o & x& x& x& x& x& x& x& x& x& x  \\ 
\bottomrule
\end{tabularx}
%\end{adjustwidth}
\end{table*}
latter served iteratively as compared to the first one which starts as an empty queue. Simulation proved, the proposed mechanism is 2 to 3 times better compared with other service placement and request scheduling algorithm. Fajardo et al. \cite{aguero_radio-aware_2015} presented a novel mobile edge scheduler named MESch that orchestrate the network traffic flows at the mobile cell where different users reside. MESch prioritizes these flows to reduce the mean flow delay. It takes the conditions of each mobile user in CG-CRR slot along with current flow information. Flow sizes are not known to the scheduler in advance but the distribution of flow sizes known. Experiments show that the proposed work is near-optimal compared with other competitors. The author would like to improve this work as future work and also eager to see the malfunction aspects of MESch.
\subsubsection{Execution Cost} One of the challenging tasks in the Edge Cloud is to optimize the cost while offloading the task into edge and cloud. In cost optimization, Service Providers (SPs) wants to earn maximum from production but Mobile User (MU) wants at the same time to process their computation at a very minimal cost. There are many service provider that provides their service using different pricing models as Amazon provides On-Demand Resource Model. We have categorized heuristic algorithms with QoS parameters that are presented in Table~\ref{heuristic_qos_one}.
\subsubsection{Energy Consumption} Computation demand is increasing day by day so is the power consumption at the Data Center (DC). According to the report, only US data centres had consumed 73 billion kWh in 2020 \cite{shehabi_united_2016}. This huge consumption also emits carbon footprints that are not Eco-friendly. The energy consumption issue becomes more challenging when there is a heterogeneous resource in the system.
\subsubsection{Fault Tolerance} Fault tolerance holds great importance in scheduling. Hardware and Software faults are now a common thing when your resources are remote. There is very little work done in the Edge Cloud where fault-tolerance is considered. We have categorized heuristic algorithms with different constraints parameters that are presented in Table~\ref{heuristic_constraints_one}.

\begin{table*}[htbp]
\caption{Summary of Heuristic Scheduling Algorithms with Constraints.\label{heuristic_constraints_one}}
%\begin{adjustwidth}{-\extralength}{0cm}
		\newcolumntype{C}{>{\centering\arraybackslash}X}
		
\begin{tabularx}{\textwidth}{p{.2\linewidth}p{.13\linewidth}p{.11\linewidth}p{.13\linewidth}p{.15\linewidth}p{.24\linewidth}}
\toprule
Year & Static & Dynamic & Deadline & Fault Tolerance & Testing Tool\\ 
\midrule
Meng et al. \cite{meng_fault-tolerant_2019} (2019)& no	&yes	&yes	&yes	&Simulation\\
Huawei et al. \cite{huawei_adaptive_nodate} (2017)&yes	&no	&no	&no	&Simulation \\
Hou et al. \cite{hou_asymptotically_2016} (2016)&no	&yes	&no	&no	&Simulation \\
Hu et al. \cite{hu_dynamic_2020} (2020) &no	&yes	&no	&no	&Simulation \\
Wang et al. \cite{wang_dynamic_2018} (2018) &no	&yes	&no	&no	&Simulation \\
Tuli et al. \cite{tuli_dynamic_2020} (2020) &no	&yes	&no	&no	&Real \\
Urgaonkar et al. \cite{urgaonkar_dynamic_2015} (2015) &no	&yes	&no	&no	&Simulation \\
Han et al. \cite{han_ondisc_2019} (2019) &no	&yes	&no	&no	&Simulation \\
Wang et al. \cite{wang_online_2017} (2017) &no	&yes	&no	&no	&Real \\
Zhong et al. \cite{zhong_age-aware_2019} (2019) &yes	&no	&no	&no	&Simulation \\
Farhadi et al. \cite{farhadi_service_2019} (2019) &yes	&no	&yes	&no	&Simulation \\
Wang et al. \cite{wang_computing_2019} (2019) &no	&yes	&yes	&no	&Simulation \\
Liu et al. \cite{liu_cooper-sched_2019} (2019) &yes	&no	&yes	&no	&Simulation \\
Li et al. \cite{li_cooperative_2021} (2021) &yes	&no	&yes	&no	&Simulation \\
Yin et al. \cite{yin_online_2017} (2017) &no	&yes	&yes	&no	&Simulation \\
Zhang et al. \cite{zhang_cost_2018} (2018) &yes	&no	&no	&no	&Simulation \\
Lee et al. \cite{lee_data_2020} (2020) &yes	&no	&no	&no	&Simulation \\
Yu et al. \cite{yu_energy-efficient_2018} (2018) &no	&yes	&yes	&no	&Simulation \\
Kaur et al. \cite{kaur_keids_2020} (2020) &yes	&no	&yes	&yes	&Simulation \\
Wang et al. \cite{wang_reconciling_2016} (2016) &yes	&no	&yes	&no	&Simulation \\
Katsalis et al. \cite{katsalis_sla-driven_2016} (2016) &yes	&no	&no	&no	&Simulation \\
Meng et al. \cite{meng_delay-sensitive_2019} (2019) &yes	&no	&no	&no	&Simulation \\
Huang et al. \cite{huang_task_2018} (2018) &yes	&no	&no	&no	&Simulation \\
Lin et al. \cite{lin_distributed_2018} (2018) &yes	&no	&no	&no	&Simulation \\
Sun et al. \cite{sun_eco-friendly_2020} (2020) &yes	&no	&no	&no	&Simulation \\
Zhang et al. \cite{zhang_hetero-edge_2019} (2019) &yes	&no	&no	&no	&Real \\
Zhao et al. \cite{zhao_load_2019} (2019) &yes	&no	&no	&no	&Simulation \\
Li et al. \cite{li_neighborhood_2021} (2021) &no	&yes	&no	&no	&Real \\
Wang et al. \cite{wang_incentive_2019} (2019) &yes	&no	&no	&no	&Simulation \\
Miao et al. \cite{miao_intelligent_2020} (2020) &no	&yes	&no	&no	&Simulation \\
Wei et al. \cite{wei_mobility-aware_2020} (2020) &yes	&no	&no	&no	&Real \\
Samanta et al. \cite{samanta_battle_2019} (2019) &yes	&no	&yes	&no	&Simulation \\
Sajnani et al. \cite{sajnani_latency_2018} (2018) &yes	&no	&no	&no	&Simulation \\
Lyu et al. \cite{lyu_optimal_2017} (2017) &yes	&no	&no	&yes	&Simulation \\
Ma et al. \cite{ma_study_2020} (2020) &yes	&no	&no	&no	&Simulation \\
Ma et al. \cite{ma_cooperative_2020} (2020) &yes	&no	&no	&no	&Simulation \\
Alkhalaileh et al. \cite{alkhalaileh_data-intensive_2020} (2020) &yes	&no	&yes	&no	&Real \\
Chen et al. \cite{chen_efficient_2016} (2016) &yes	&no	&no	&no	&Simulation \\
He et al. \cite{he_its_2018} (2018) &yes	&no	&no	&no	&Simulation \\
Fajardo et al. \cite{aguero_radio-aware_2015} (2015) &no	&yes	&no	&no	&Simulation \\ 
\bottomrule
\end{tabularx}
%\end{adjustwidth}
\end{table*}

\begin{table*}[htbp]
\caption{Summary of Meta-Heuristic Scheduling Algorithms with Constraints.\label{metaheuristic_constraints_one}}
%\begin{adjustwidth}{-\extralength}{0cm}
		\newcolumntype{C}{>{\centering\arraybackslash}X}
		
\begin{tabularx}{\textwidth}{p{.2\linewidth}p{.13\linewidth}p{.11\linewidth}p{.13\linewidth}p{.15\linewidth}p{.24\linewidth}}
\toprule
Year & Static & Dynamic & Deadline & Fault Tolerance & Testing tool\\ 
\midrule
Xie et al. \cite{xie_novel_2019} (2019) & yes	&no	&no	&no	&Real \\
Li et al. \cite{li_efficient_2020} (2020) & no&	yes&	yes&	no&	Simulation \\
Wang et al. \cite{wang_task_2020} (2020) & yes&	no	&yes	&no	&Simulation \\
\bottomrule
\end{tabularx}
%\end{adjustwidth}
\end{table*}

\subsection{Meta-Heuristic Algorithms}
Meta-heuristic algorithms cover a broader range of problems unlike heuristic algorithms. These are not for a specific problem and find solutions near to optimal. Here, we have categorized the meta-heuristic scheduling schemes into PSO Based Approach \cite{xie_novel_2019}, Artificial Fish Swarm Based Approach \cite{li_efficient_2020}, and Genetic Algorithm Based Approach \cite{wang_task_2020}.
\subsubsection{PSO Based Approach} Particle swarm optimization is a widely used approach in meta-heuristic algorithms developed by Kennedy and Eberhart in 1995 \cite{kennedy_particle_1995}. It is a nature-inspired optimization technique based on the behaviour of birds. PSO usually take care of NP-hard issues in scheduling and resource provisioning. The following work has implemented the PSO based strategies. Xie et al. \cite{xie_novel_2019} proposes a technique namely Novel Directional and Non-local-Convergent Particle Swarm Optimization (DNCPSO) based on Particle Swarm Optimization (PSO) algorithm. It is based on three steps in first, non-linear inertia weight is applied to balance search capabilities. Then in the second phase, to make the particle search more directional position along with the velocity is updated. In the last stage, selection and mutation operations have been applied to keep the diversity of the particles population. Simulations depicted that, DNCPSO dramatically improves the makespan and cost as compared to baseline algorithms.
\subsubsection{Artificial Fish Swarm Based Approach} In swarm-based intelligence algorithms, the artificial fish swarm algorithm \cite{neshat_artificial_2014} is the promising optimization approach based on the behaviour of the fish since they prefer to live in colonies and likewise manifest intelligent behaviour. The following study depicted the Artificial Fish Swarm Algorithm (AFSA) as given below. Li et al. \cite{li_efficient_2020} proposed a two-level scheduling strategy for the tasks. At first-level scheduling is based on artificial fist swarm-based scheduling where most of the tasks will be scheduled first-hand. At second-level scheduling, the remaining tasks will be divided into equal size and scheduled in the main centralized data centre by considering the load balancing factor into account. Simulation results proved that the proposed work is efficient in term of job completion time and total cost.

\subsubsection{Genetic Algorithm Based Approach} Another significant optimization approach is the Genetic Algorithm (GA), based on the evolution concept and was invented by John Holland in 1960 \cite{jong_analysis_1975,holland_adaptation_1992}. GA has been tested in many optimization problems and depicted an acceptable rate of success. In GA, there is a set of possible solutions for a problem, then it goes through crossovers and mutations like in natural genetics, producing a novel successor. This process is iterative and runs until finding an acceptable solution. The following work depicted the GA based scheduling strategies. Wang et al. \cite{wang_task_2020} discusses a technique that is based on Genetic Algorithm (GA). The penalty function is used for the determination of time execution based on delays. To achieve the global optimum, the proposed Catastrophic Genetic Algorithm (CGA) also incorporate the cataclysm in it. The result shows that the algorithm outperforms in term of completion time shortening based on task delays. Summaries of surveyed meta-heuristic scheduling algorithms along with advantages, demerits, QoS and Constraints are presented in Table~\ref{metaheuristic_advantages_one}, \ref{metaheuristic_qos_one} and \ref{metaheuristic_constraints_one}.
\section{Future directions}\label{tab:future}
There are still many problems in scheduling algorithms like heterogeneity, complex nature of task from the mobile users, energy consumption, mobile users unpredictable requirements like computation, storage and memory. Only one scheduling algorithm cannot satisfy all the requirement at a time. Some focus on availability and reliability, but others ignore these and focus on cost and energy etc. Therefore, future scheduling algorithm should incorporate all these limitations.
In this section, we have reviewed the future direction and issues not examined by the researchers.
\begin{itemize}
    \item Scheduling decision is a dominant stage that determines where to offload the task/workload either at the local device, Edge Cloud or cooperatively at both bearings.
    \item A plethora of papers considers only static scenarios of scheduling. Hence, more sophisticated strategies such as employing prediction techniques respecting mobility and channel quality to gauge the scheduling cost under diverse schemas are essential.
    \item Most of the algorithms ignore the network bandwidth concept that could lead to network failure and increase delays.
    \item To date, nearly all papers supposed the Edge Cloud architecture levelled in a sense that computes nodes are identical in terms of power (computing) and distribution. However, it would be a point of interest to weigh up the distinct hierarchical structures of computing resources in the Edge Cloud architecture.
    \item Besides, proposed scheduling approaches and implementations are demonstrated based on numerical analysis and simulations. Further, many of them took a very naive and even unfeasible adoption. Hence, it is necessary to consider more practical and realistic scenarios as adopted by \cite{tuli_dynamic_2020,wei_mobility-aware_2020,alkhalaileh_data-intensive_2020}since they evaluate the real-world traces. Moreover, a real network context is imperative to demystify the alpha and the omega of the Edge Cloud collaboration.
\end{itemize}
\section{Conclusion}\label{tab:Conclusion}
In this paper, we have reviewed the scheduling approaches in the Edge Cloud domain. The core reason behind the scheduling algorithm is to schedule a workload/task at the intended edge or cloud location to improve the multiple key performance parameters. Categorization of the reviewed papers based on the algorithm used, nature of the task (independent or workflow), the technique used, advantages, limitations along with QoS parameters (e.g., execution time,  makespan time, execution cost, response time, resource utilization, energy consumption, throughput, SLA violations, availability, scalability) and profit in connection with different constraints (e.g., mobility (static/dynamic), fault tolerance and testing tool) used have presented in detail. Besides, machine learning-based papers have also been presented explicitly and categorized into supervised learning, unsupervised learning and reinforcement learning based on dominance and demerits. In machine learning-based approaches, many papers only consider delay and energy as the objective function. Therefore, it is required to incorporate more parameters. Simulation tools have also discussed either the proposed scheduling technique is validated in a real-world environment or simulation. The main objective behind this paper is to give a broader and deeper understanding regarding the scheduling approaches in the Edge Cloud environment that paves the way in the invention of novel and state of the art scheduling approaches. Overall we investigated 50 research papers where scheduling is the main focus in the Edge Cloud environment. The Edge Cloud is still in its infancy stage and a lot of efforts needed in this regard. We conjecture that the presented survey will prove a stepping stone into the Edge Cloud scenario and equally beneficial for the academic and industrial arena.

\bibliographystyle{IEEEtran}
\bibliography{SchedulingAtEdge}

\end{document}